\documentclass{revtex4}
\usepackage{amssymb}

\usepackage{amsfonts}
\usepackage{amsmath}

\input{tcilatex}

\begin{document}

\preprint{}
\title{ $2+1-$dimensional electrically charged black holes in Einstein -
Power - Maxwell Theory}
\author{O. Gurtug}
\email{ozay.gurtug@emu.edu.tr}
\author{S. Habib Mazharimousavi}
\email{habib.mazhari@emu.edu.tr}
\author{M. Halilsoy}
\email{mustafa.halilsoy@emu.edu.tr}
\affiliation{Department of Physics, Eastern Mediterranean University, G. Magusa, north
Cyprus, Mersin 10 - Turkey.}
\keywords{Black-holes, Power Maxwell Fields, 3D Gravity, Non-linear
electrodynamics, exact solution, Einstein-Maxwell theory}
\pacs{PACS number}

\begin{abstract}
A large family of new black hole solutions in $2+1-$dimensional
Einstein-Power-Maxwell (EPM) gravity with prescribed physical properties is
derived. We show with particular examples that according to the power
parameter $k$ of the Maxwell field, the obtained solutions may be
asymptotically flat for $1/2<k<1$ or non-flat for $k>1$ in the vanishing
cosmological constant limit. We study the thermodynamic properties of the
solution with two different models and it is shown that thermodynamic
quantities satisfy the first law. The behaviour of the heat capacity
indicates that by employing the $1+1-$dimensional dilaton analogy the local
thermodynamic stability is satisfied.
\end{abstract}

\maketitle

\section{Introduction}

It is well known that the main motivation to study nonlinear electrodynamics
(NED) was to overcome some of the difficulties that occur in the standard
linear Maxwell theory. Divergence in self-energy due to point charges was
one such difficulty that kept physics community busy for decades. The model
of Born - Infeld NED was developed with the hope to remove these
divergences; see e.g.\cite{1,2,3,4,5,6}, and by a similar trend NED was
employed to eliminate black hole singularities in general theory of
relativity. A striking example of regular black hole solutions in $\left(
3+1\right) -$dimensions was given in \cite{7} that considers Einstein field
equations coupled with NED which satisfies the weak energy condition and
recovers the Maxwell theory in the weak field limit. The use of NED for
eliminating singularities was also proved successful in $\left( 2+1\right) -$
dimensions \cite{8}.

During the last decade, $(2+1)$ - dimensional spacetimes admitting black
hole solutions have attracted much attention. As a matter of fact the $d=3$
case singles out among other $(d\geq 4)$ spacetimes with special mass and
charge dependence. The first example in this regard is the BTZ black hole 
\cite{9}. Later on, Einstein-Maxwell \cite{10} and Einstein-Maxwell-dilaton 
\cite{11} extensions have also been found. The black hole solutions found in
this context include all typical characteristics that can be found in $(3+1)$
or higher dimensional black holes such as; horizon(s), black hole
thermodynamics and Hawking radiation. The black hole solution derived in 
\cite{12} is another example for $(2+1)-$dimensions within the context of a
restricted class of NED in which the Maxwell scalar has a power in the form
of $\left( F_{\mu \nu }F^{\mu \nu }\right) ^{3/4}$. This particular power
results from imposing traceless condition on the energy-momentum tensor.

The main objective of the present study is two-fold. First, we construct a
large class of black hole solutions sourced by the power Maxwell field in
which the Maxwell scalar has the form $\left( F_{\mu \nu }F^{\mu \nu
}\right) ^{k}$. Here, the power $k$ is \ a real rational number which will
be restricted to some intervals as a requirement of the energy conditions.
In general, for $d$ - dimensional spacetimes, the specific choice of $k=%
\frac{d}{4}$ yields a traceless Maxwell's energy momentum tensor \cite{13}
which is known to satisfy the conformal invariance condition. In recent
years, the use of power Maxwell fields has attracted considerable interest.
It has been used for obtaining solutions in $d-$ spacetime dimensions \cite%
{14}, Ricci flat rotating black branes with a conformally Maxwell source 
\cite{15}, Lovelock black holes \cite{16}, Gauss-Bonnet gravity \cite{17}
and the effect of power Maxwell field on the magnetic solutions in
Gausss-Bonnet gravity \cite{18}. Therefore in \cite{12}, the power $3/4$ of
the Maxwell scalar in $\left( 2+1\right) -$dimensions is the unique case
which results from this traceless condition. Our first motivation in this
study is to find the most general solution in $(2+1)-$dimensional
Einstein-Power-Maxwell (EPM) spacetime without imposing the traceless
condition. Stated otherwise, choosing a traceful energy momentum tensor
amounts to treating $k$ as a new parameter and we wish to investigate this
freedom as much as we can. However, our analysis on the obtained solutions
has revealed that the power parameter $k$ can not be arbitrary. For a
physically acceptable solution it must be a rational number. Hence, our
general solution overlaps with the solution presented in \cite{12}, if one
takes the power parameter $k=\frac{3}{4}.$ Depending on the value of $k$,
however, the resulting metric displays different characteristics near $r=0$
which makes the present study more interesting. With the freedom of $k$ we
explore a rich possibility in the structure of singularities. For values $%
1/2<k<1,$ for instance, the resulting spacetime becomes asymptotically flat
in the vanishing cosmological constant $(\Lambda =0)$, \ and for $\Lambda >0,
$ it is the asymptotically de-Sitter spacetime. For $k>1$ the resulting
spacetime is non-asymptotically flat. Furthermore, the resulting metric
depends not only on the parameter $k$ but also on the mass $M$, the charge $Q
$ and the cosmological constant $\Lambda .$ When $\Lambda >0,$ the solution
describes a charged de-Sitter black hole spacetime with inner and outer
horizons for $1/2<k<1,$ and a cosmological horizon for $k>1$. For specific
values of these parameters the resulting spacetime singularity at $r=0$ is
naked whose strength becomes $k$ dependent. When the cosmological constant $%
\Lambda <0,$ the resulting spacetime corresponds to charged anti de-Sitter
with a cosmological horizon in the range of the power $1/2<k<1.$ For $k>1$,
the resulting charged de-Sitter spacetime becomes naked singular at $r=0$.

Another important issue in black hole physics is the concept and analysis of
thermodynamic properties. This issue has gained a significant momentum not
just in the linear Maxwell theory but also in the NED. As an example in \cite%
{19}, higher dimensional gravity coupled to NED sourced by power Maxwell
field has been analyzed thermodynamically for $d>3$. The local and global
thermodynamic stability is investigated by calculating the Euclidean action
with appropriate boundary term in the grand canonical ensemble. Our second
objective in this study is to investigate the local thermodynamic stability
of the resulting black holes. This is achieved by employing the method
presented in \cite{20}, in which the local Hawking temperature is found from
the Unruh effect. We calculate the heat capacities and show that our
solution conditionally displays local thermodynamic stability. For specific
values of the parameters, the calculated specific heat capacity at constant
charge and electric potential both change sign at particular points. This
behaviour indicates that there may be a possible phase change in the black
hole state. Alternatively, for a thorough thermodynamical analysis we appeal
to the dilatonic analogy established in $1+1-$dimensions \cite{25,26,27}.

The organization of the paper is as follows. Section II, introduces the
theory of EPM with solution and spacetime structure. Thermodynamic
properties of the solution is considered in Section III. We complete the
paper with Conclusion in Section IV.

\section{Einstein- Power - Maxwell (EPM) Solutions and Spacetime Structure}

The $3-$dimensional action for EPM theory with cosmological constant $%
\Lambda $ is given by ($c=k_{B}=\hslash =8G=1$)

\begin{equation}
\begin{tabular}{l}
$I=\int dx^{3}\sqrt{-g}\left( \frac{1}{2\pi }\left( R-\frac{2}{3}\Lambda
\right) -L\left( \mathcal{F}\right) \right) $%
\end{tabular}%
,
\end{equation}%
in which $L\left( \mathcal{F}\right) =\left\vert \mathcal{F}\right\vert ^{k}$
and $\mathcal{F}$ is the Maxwell invariant 
\begin{equation}
\mathcal{F}=F_{\mu \nu }F^{\mu \nu },\text{ }  \notag
\end{equation}%
while the parameter $k$ is arbitrary for the time being. Variation with
respect to the gauge potential $\mathbf{A}$ yields the Maxwell equations 
\begin{equation}
\mathbf{d}\left( ^{\star }\mathbf{F}L_{\mathcal{F}}\right) =0\rightarrow 
\mathbf{d}\left( ^{\star }\mathbf{F}\left\vert \mathcal{F}\right\vert
^{k-1}\right) =0,
\end{equation}%
where $^{\star }$ denotes duality. Variation of the action with respect to
the spacetime metric $g_{\mu \nu }$ yields the field equations

\begin{equation}
G_{\mu }^{\nu }+\frac{1}{3}\Lambda \delta _{\mu }^{\nu }=\pi T_{\mu }^{\nu },
\end{equation}%
where 
\begin{equation}
T_{\ \nu }^{\mu }=\frac{1}{2}\left( 4\left( F_{\nu \lambda }F^{\ \mu \lambda
}\right) L_{\mathcal{F}}-\delta _{\ \nu }^{\mu }L\right) ,
\end{equation}%
is the energy-momentum tensor of the power Maxwell field and explicitly reads%
\begin{equation}
T_{\ \nu }^{\mu }=\frac{\left\vert \mathcal{F}\right\vert ^{k}}{2}\left( 
\frac{4k\left( F_{\nu \lambda }F^{\ \mu \lambda }\right) }{\mathcal{F}}%
-\delta _{\ \nu }^{\mu }\right) .
\end{equation}%
Our metric ansatz for $(2+1)-$ dimensions, is chosen as 
\begin{equation}
ds^{2}=-f\left( r\right) dt^{2}+\frac{dr^{2}}{f\left( r\right) }%
+r^{2}d\theta ^{2}.
\end{equation}%
Static, electrically charged potential ansatz is given by%
\begin{equation*}
\mathbf{A}=A(r)dt,\text{ \ \ }
\end{equation*}%
which leads to%
\begin{equation}
\mathbf{F=dA=}E(r)dr\wedge dt,
\end{equation}%
with its dual 
\begin{equation}
^{\star }\mathbf{F=}E(r)rd\theta 
\end{equation}%
and%
\begin{equation}
\mathcal{F}=F_{\mu \nu }F^{\mu \nu }=-2E(r)^{2}.
\end{equation}%
Accordingly, the Maxwell equation reads now%
\begin{equation}
\mathbf{d}\left( E(r)rd\theta \left[ 2E(r)^{2}\right] ^{k-1}\right) =0,
\end{equation}%
which leads to the solution as%
\begin{equation}
rE(r)^{2k-1}=\text{constant},
\end{equation}%
or equivalently%
\begin{equation}
E(r)=\frac{\text{constant}}{r^{\frac{1}{2k-1}}}.
\end{equation}%
By using the latter result in (7) and choosing the integration constant
proportional to the electric charge $Q,$ one obtains the potential%
\begin{equation}
A(r)=\left\{ 
\begin{array}{cc}
Q\ln r & k=1 \\ 
\frac{Q\left( 2k-1\right) }{2\left( k-1\right) }r^{\frac{2\left( k-1\right) 
}{\left( 2k-1\right) }} & k\neq 1,\frac{1}{2}%
\end{array}%
\right. .
\end{equation}%
The resulting energy - momentum tensor follows from (5) as,%
\begin{equation}
T_{\ \nu }^{\mu }=\frac{1}{2}\left\vert \mathcal{F}\right\vert ^{k}\text{diag%
}\left( \xi ,\xi ,-1\right) 
\end{equation}%
where $\xi =\left( 2k-1\right) $ and the explicit form of $\mathcal{F}$ is
given by 
\begin{equation}
\mathcal{F}=-\frac{Q^{2}}{r^{\frac{2}{2k-1}}}
\end{equation}%
in which we recall that $Q$ is a constant related to the charge of the black
hole. One can show that the WEC and SEC restrict us to the set $k\in \left( 
\frac{1}{2},\infty \right) $ (see Appendix). The $tt$ component of Einstein
equations (3) reads%
\begin{equation}
\frac{1}{2r}f^{\prime }\left( r\right) +\frac{1}{3}\Lambda =4\pi \xi
\left\vert \mathcal{F}\right\vert ^{k}.
\end{equation}%
whose integration gives,%
\begin{equation}
f\left( r\right) =D+\frac{r^{2}}{l^{2}}-\frac{\pi \left( 2k-1\right) ^{2}}{%
2\left( k-1\right) }Q^{2k}r^{\frac{2\left( k-1\right) }{2k-1}},
\end{equation}%
in which $D$ is an integration constant, $\Lambda =-1/l^{2}$ and $k\neq 1.$
The choice $k=1$, which we exclude here gives the known charged BTZ black
hole solution in Einstein-Maxwell theory \cite{10}. We note that for $k=1$, $%
\mathcal{F}=-\frac{Q^{2}}{r^{2}}$ diverges at $r=0$ which is weaker than the
cases for $k<1$. This behaviour, however, turns opposite for the choice $k>1$%
. In order to illustrate this important effect of the power parameter $k$,
we calculate the Kretschmann invariant for $k=1$, $k<1$ and $k>1$. Since the
resulting expressions for any $k>1$ or $k<1$ is too complicated, we prefer
to calculate the Kretschmann invariant for specific values of $k$;%
\begin{eqnarray}
\mathcal{K} &\mathcal{=}&\frac{12}{l^{4}}-\frac{4Q^{2}}{r^{2}}\left( \frac{2%
}{l^{2}}-\frac{Q^{2}}{r^{2}}\right) \text{ \ \ \ \ \ \ \ \ \ \ \ for \ \ }\
k=1, \\
\mathcal{K} &\mathcal{=}&\frac{12}{l^{4}}-\frac{8Q^{2}}{r^{6}}\text{ \ \ \ \
\ \ \ \ \ \ \ \ \ \ \ \ \ \ \ \ \ \ \ \ \ \ \ \ \ \ \ for \ \ }\ k=3/4, 
\notag \\
\mathcal{K} &=&\frac{12}{l^{4}}-\frac{2\pi Q^{4}}{r^{4/3}}\left( \frac{10}{%
l^{2}}-\frac{19\pi Q^{4}}{2r^{4/3}}\right) \text{ \ \ \ \ \ \ for }k=2. 
\notag
\end{eqnarray}%
It is clear from these results that the rate of divergence of the
Kretschmann invariant for $k=\frac{3}{4}$ is faster than the cases $k\geq 1.$%
The case when the power of $r,$ $\frac{2\left( k-1\right) }{2k-1}<0,$ bounds
the value of $k$ to $\frac{1}{2}<k<1$ which is also consistent with the
energy conditions. Note that this case corresponds to asymptotically flat
spacetime if one takes $\Lambda =0.$ 

The integration constant $D$ can be associated with the mass of the black
hole i.e., $D$ can be expressed in terms of mass at infinity by employing
the Brown-York \cite{12,21,22} formalism. Following the quasilocal mass
formalism it is known that, a spherically symmetric $3-$dimensional metric
solution as%
\begin{equation}
ds^{2}=-F(r)^{2}dt^{2}+\frac{1}{G(r)^{2}}dr^{2}+r^{2}d\theta ^{2}
\end{equation}%
admits a quasilocal mass $M_{QL}$ defined by 
\begin{equation}
M_{QL}=\lim_{r_{b}\rightarrow \infty }2F(r_{b})\left[ G_{r}(r_{b})-G(r_{b})%
\right] .
\end{equation}%
Here $G_{r}(r_{b})$ is an arbitrary non-negative reference function, which
yields the zero of the energy for the background spacetime, and $r_{b}$ is
the radius of the space-like hypersurface. According to our line element we
get%
\begin{eqnarray}
F(r)^{2} &=&G(r)^{2}=D+\frac{r^{2}}{l^{2}}-\frac{\pi \left( 2k-1\right) ^{2}%
}{2\left( k-1\right) }Q^{2k}r^{\frac{2\left( k-1\right) }{2k-1}}, \\
G_{r}(r)^{2} &=&\frac{r^{2}}{l^{2}}-\frac{\pi \left( 2k-1\right) ^{2}}{%
2\left( k-1\right) }Q^{2k}r^{\frac{2\left( k-1\right) }{2k-1}}
\end{eqnarray}%
which yield%
\begin{equation}
M_{QL}=\lim_{r_{b}\rightarrow \infty }2\left( \frac{r_{b}^{2}}{l^{2}}\left(
1+\frac{l^{2}}{2r_{b}^{2}}D-\frac{l^{2}ar_{b}^{\frac{2\left( k-1\right) }{%
2k-1}}}{2r_{b}^{2}}\right) -(D+\frac{r_{b}^{2}}{l^{2}}-ar_{b}^{\frac{2\left(
k-1\right) }{2k-1}})\right) .
\end{equation}%
Here we expanded the square roots and $a=\frac{\pi \left( 2k-1\right) ^{2}}{%
2\left( k-1\right) }Q^{2k}.$ We note that since $\frac{2\left( k-1\right) }{%
2k-1}<1$ for all values of $1/2<k<\infty ,$ this limit results in $-D,$
independent of the value of $k.$ We would like to add that the same result
may be found by applying the method introduced in \cite{23,24} with a proper
choice of the background metric.

Therefore the metric function, irrespective of the power of $r,$ for $M>0$
is given by

\begin{equation}
f\left( r\right) =-M+\frac{r^{2}}{l^{2}}-\frac{\pi \left( 2k-1\right) ^{2}}{%
2\left( k-1\right) }Q^{2k}r^{\frac{2\left( k-1\right) }{2k-1}}.
\end{equation}%
Finally in this section we give the Ricci scalar 

\begin{equation}
R=-\frac{6}{l^{2}}+\frac{\pi Q^{2k}\left( 4k-3\right) }{r^{\frac{2k}{2k-1}}}.
\end{equation}%
which indicates the occurrence of true curvature singularity for any $k>%
\frac{1}{2}$. Although the particular choice $k=\frac{3}{4}$ shows $R$ to be
regular at $r=0$ this is not supported by the Kretschmann scalar expression
in (18). Nevertheless, the energy conditions (i.e. at least WEC and SEC) -
given in Appendix - always result in negative exponents in radial coordinate 
$r$ therefore $r=0$ is a true curvature singularity.

\section{Thermodynamics}

\subsection{Analysis with finite boundary model}

In this section, we study the thermodynamical properties of the solution
(24). A similar analysis for $d-$dimensional charged black holes with a NED
sourced by power Maxwell fields was considered in \cite{11}, by employing
Euclidean action with a suitable boundary term in the grand canonical
ensemble. The analysis was carried out for spacetime dimensions $d>3$.

In this study, we follow an alternative method as demonstrated in \cite{19}
to find the local Hawking temperature by using the Unruh effect in curved
spacetime which is equivalent to finding the periodicity in the time
coordinate in the Euclidean version of the metric covering the outer region
of the black hole. In the Unruh effect, an observer outside the black hole
experiences a thermal state with local temperature defined by

\begin{equation}
T_{H}(r)=\frac{2f^{\prime }(r_{h})}{\pi \sqrt{-\chi _{\alpha }\chi ^{\alpha }%
}}=\frac{32}{\pi \sqrt{f(r)}}\left\{ \frac{r_{h}}{l^{2}}-\frac{\pi \left(
2k-1\right) Q^{2k}}{2r_{h}^{\frac{1}{2k-1}}}\right\} ,
\end{equation}%
where $\chi ^{\alpha }$ is the Killing vector field generating the outer
horizon and \ the location of the horizons are given by the roots of $%
f(r_{h})=0$ which implies

\begin{equation}
M=\frac{r_{h}^{2}}{l^{2}}-\frac{\pi \left( 2k-1\right) ^{2}}{2\left(
k-1\right) }Q^{2k}r_{h}^{\frac{2\left( k-1\right) }{2k-1}}.
\end{equation}%
It should be noted that the power parameter $k$ in the analysis of
thermodynamic properties is assumed to satisfy $1/2<k<1.$ It is remarkable
to note that in the limits, $T_{H}(r)\mid _{r\rightarrow r_{h}}\rightarrow
\infty $ and $T_{H}(r)\mid _{r\rightarrow \infty }\rightarrow 0.$ This is
expected because the solution given in (24) is non asymptotically flat,
hence, we have a vanishing temperature (i.e. from (26)) at infinity.
Following the same procedure as demonstrated in \cite{19}, we define the
reenergized temperature as $T_{\infty }=T_{H}(r)\sqrt{-\chi _{\alpha }\chi
^{\alpha }}=\frac{f^{\prime }(r_{h})}{4\pi }$ which gives

\begin{equation}
T_{\infty }=\frac{4}{\pi }\left\{ \frac{r_{h}}{l^{2}}-\frac{\pi \left(
2k-1\right) Q^{2k}}{2r_{h}^{\frac{1}{2k-1}}}\right\} .
\end{equation}

The internal energy of the system on a constant $t$ hypersurface can be
defined from the Brown-York \cite{12,21,22} quasilocal energy formalism as

\begin{equation}
E(r_{b})=-2\left( \sqrt{f\left( r_{b}\right) }-\frac{r_{b}}{l}\right) 
\end{equation}%
where $r=r_{b}$ is a finite boundary of the spacetime. One may vary the
internal energy $E(r_{b})$ with respect to $r_{h}$ and $Q$ which leads to
the first law of thermodynamics in the following form

\begin{equation}
dE(r_{b})=T_{H}(r_{b})dS+\Phi \left( r_{b}\right) de,
\end{equation}%
in which the entropy $S,$ after our unit convention ($c=k_{B}=\hslash =8G=1$%
) is given by

\begin{equation}
S=4\pi r_{h},
\end{equation}%
and $e$ is the electric charge

\begin{equation}
e=\frac{\sqrt{2}}{4}\pi k\left( 2k-1\right) ^{\frac{2k-1}{2k}}Q^{2k-1}.
\end{equation}%
Here, 
\begin{equation}
T_{H}(r_{b})=\frac{1}{2\pi \sqrt{f\left( r_{b}\right) }}\left\{ \frac{r_{h}}{%
l^{2}}-\frac{\pi }{2}\left( 2k-1\right) Q^{2k}r_{h}^{-\frac{1}{2k-1}%
}\right\} 
\end{equation}%
is the Hawking temperature at the boundary of the black hole spacetime and%
\begin{equation}
\Phi \left( r_{b}\right) =\frac{2\left( 2k-1\right) ^{\frac{1}{2k}}\sqrt{2}Q%
}{\left( 1-k\right) \sqrt{f\left( r_{b}\right) }}\left( r_{h}^{\frac{2\left(
k-1\right) }{2k-1}}-r_{b}^{\frac{2\left( k-1\right) }{2k-1}}\right) 
\end{equation}%
is the electric potential difference between the horizon and boundary $r_{b}$%
. On the other hand the electric potential difference between the boundary
and infinity is given by%
\begin{equation}
\Psi \left( r_{b}\right) =\frac{2\left( 2k-1\right) ^{\frac{1}{2k}}\sqrt{2}Q%
}{\left( 1-k\right) \sqrt{f\left( r_{b}\right) }}\left( r_{b}^{\frac{2\left(
k-1\right) }{2k-1}}\right) .
\end{equation}

We consider the black hole inside a box bounded by $r=r_{b}$ and calculate
the heat capacity of the black hole at constant $Q$, $\Phi $ and $\Psi .$
The local thermodynamical stability conditions are determined by the sign of
the heat capacities calculated at constant quantities in the limit of large
values of $r_{b}.$ which is defined by,%
\begin{equation}
C_{X}\equiv T\left( \frac{\partial S}{\partial T}\right) _{X}\geq 0,\text{ \
\ }
\end{equation}%
in which $T=T_{H}\left( r_{b}\right) $ and $X$ is the quantity to be held
constant$.$ Note that, we consider $S=S(r_{h})$ and $T=T(r_{h},Q^{2k})$,
therefore Eq. (36) can be written as%
\begin{equation}
C_{X}\equiv T\left( \frac{\partial S}{\partial T}\right) _{X}=T\left( \frac{%
\partial T}{\partial S}\right) _{X}^{-1}=T\left\{ \left( \frac{\partial T}{%
\partial r_{h}}\right) \left( \frac{\partial r_{h}}{\partial S}\right)
+\left( \frac{\partial T}{\partial Q^{2k}}\right) \left( \frac{\partial
Q^{2k}}{\partial S}\right) _{X}\right\} ^{-1},
\end{equation}%
and with $\frac{\partial S}{\partial r_{h}}=4\pi ,$ latter equation reduces
to,%
\begin{equation}
C_{X}=\frac{4\pi T}{\left\{ \left( \frac{\partial T}{\partial r_{h}}\right)
+\left( \frac{\partial T}{\partial Q^{2k}}\right) \left( \frac{\partial
Q^{2k}}{\partial r_{h}}\right) _{X}\right\} }.
\end{equation}%
The heat capacities for constant $Q$, $\Phi $ and $\Psi $ are calculated by
using Eq. (38). Because of the complexity of the resulting expressions we
prefer to give only the expressions for large values of $r_{b},$ hence, the
limiting heat capacities as $r_{b}\rightarrow \infty $ are

\begin{eqnarray}
C_{Q} &=&T\left( \frac{\partial S}{\partial T}\right) _{Q}\simeq \frac{4\pi
r_{h}\left[ r_{h}^{\frac{2k}{2k-1}}-\frac{\pi l^{2}Q^{2k}\left( 2k-1\right) 
}{2}\right] }{\left( r_{h}^{\frac{2k}{2k-1}}+\frac{\pi l^{2}Q^{2k}}{2}%
\right) }, \\
C_{\Phi } &=&T\left( \frac{\partial S}{\partial T}\right) _{\Phi }\simeq 
\frac{4\pi r_{h}\left[ r_{h}^{\frac{2k}{2k-1}}-\frac{\pi l^{2}Q^{2k}\left(
2k-1\right) }{2}\right] }{\left( r_{h}^{\frac{2k}{2k-1}}+\pi
l^{2}Q^{2k}\left( 2k-1\right) ^{2}\right) },  \notag \\
C_{\Psi } &=&T\left( \frac{\partial S}{\partial T}\right) _{\Psi }\simeq 
\frac{4\pi r_{h}\left[ r_{h}^{\frac{2k}{2k-1}}-\frac{\pi l^{2}Q^{2k}\left(
2k-1\right) }{2}\right] }{\left( r_{h}^{\frac{2k}{2k-1}}+\frac{\pi
l^{2}Q^{2k}}{2}\right) }.  \notag
\end{eqnarray}%
One observes that since, from (28),$\ f^{\prime }\left( r_{h}\right) >0$ i.e.

\begin{equation}
r_{h}^{\frac{2k}{2k-1}}-\frac{\pi l^{2}Q^{2k}\left( 2k-1\right) }{2}>0,
\end{equation}%
thermodynamically our solution indicates a locally stable black hole.

\subsection{Analysis with 1+1-dimensional dilaton gravity model}

In this section, we employ the method presented in \cite{25,26,27} to study
the thermodynamics of the EPM black hole found above. In this method the
solution given in Eq. (24) will be obtained from the dilaton and its
potential of two-dimensional dilaton gravity through dimensional reduction.
Now we consider 
\begin{equation}
ds^{2}=g_{ab}dx^{a}dx^{b}=\tilde{g}_{\mu \nu }d\tilde{x}^{\mu }d\tilde{x}%
^{\nu }+\phi ^{2}\left( \tilde{x}\right) d\theta ^{2}
\end{equation}%
where $\phi $ denotes the radius of the circle $S^{1}$ in $M_{3}=M_{2}\times
S^{1}.$ The Greek indices represent the two-dimensional spacetime. After the
Kaluza-Klein dimensional reduction, the action (1) reads as,%
\begin{equation}
S_{2D}=2\pi \int d\tilde{x}^{2}\sqrt{-\tilde{g}}\phi \left( \frac{\tilde{R}-2%
\tilde{\Lambda}}{2\pi }-L\left( \mathcal{F}\right) \right) ,\text{ \ \ \ \ }%
L\left( \mathcal{F}\right) =\left\vert \mathcal{F}\right\vert ^{k},
\end{equation}%
in which $\tilde{R}$ is the Ricci scalar of $M_{2}$ and $\tilde{\Lambda}%
=\Lambda /3.$ Varying the above action leads to the following field equations%
\begin{eqnarray}
d\left( \phi L_{\mathcal{F}}\ {}^{\star }\mathbf{F}\right)  &=&0, \\
\nabla ^{2}\phi +2\phi \tilde{\Lambda} &=&2\pi \phi \left( L\left( \mathcal{F%
}\right) -2\mathcal{F}L_{\mathcal{F}}\right) , \\
\tilde{R}-2\tilde{\Lambda} &=&-2\pi L\left( \mathcal{F}\right) .
\end{eqnarray}%
Herein, the electric field $2-$form is given by $\mathbf{F=}E\left( \phi
\right) dt\wedge d\phi $ and its dual ${}$becomes $0-$form $^{\star }\mathbf{%
F=}E\left( \phi \right) .$ Note that, $\phi $ is effectively one of our
coordinates. The electric field invariant $F_{ab}F^{ab}$ is%
\begin{equation}
\mathcal{F}=-\frac{1}{2}E\left( \phi \right) ^{2}
\end{equation}%
which implies from (43)%
\begin{equation}
E\phi \left( E^{2}\right) ^{k-1}=\text{constant}.
\end{equation}%
The latter equation yields the following electric field%
\begin{equation}
E(\phi )=\frac{q}{\phi ^{\frac{1}{2k-1}}},
\end{equation}%
where $q$ is an integration constant. Then, the Lagrangian $\mathcal{L}%
\left( F\right) $ can be written as, 
\begin{equation}
L\left( \mathcal{F}\right) =\frac{1}{2^{k}}\frac{q^{2k}}{\phi ^{\frac{2k}{%
2k-1}}},
\end{equation}%
and 
\begin{equation}
L_{\mathcal{F}}=\frac{-k}{2^{\left( k-1\right) }}\frac{q^{2\left( k-1\right)
}}{\phi ^{\frac{2\left( k-1\right) }{2k-1}}}.
\end{equation}%
The rest of the field equations are given by 
\begin{equation}
\nabla ^{2}\phi =V\left( \phi \right) =-2\phi \tilde{\Lambda}+2\pi \phi
\left( \frac{1}{2^{k}}\frac{q^{2k}}{\phi ^{\frac{2k}{2k-1}}}\right) \left(
1-2k\right) 
\end{equation}%
and 
\begin{equation}
\tilde{R}=-V^{\prime }\left( \phi \right) =2\tilde{\Lambda}-2\pi \left( 
\frac{1}{2^{k}}\frac{q^{2k}}{\phi ^{\frac{2k}{2k-1}}}\right) .
\end{equation}%
It is remarkable to observe that, these equations correspond to the $2-$
dimensional field equations of dilaton gravity with an action%
\begin{equation}
S_{2D}=\int_{M_{2}}d\tilde{x}dt\sqrt{-\tilde{g}}\left( \phi \tilde{R}%
+V\left( \phi \right) \right) 
\end{equation}%
and the line element 
\begin{equation}
ds^{2}=-f\left( \tilde{x}\right) dt^{2}+\frac{d\tilde{x}^{2}}{f\left( \tilde{%
x}\right) }.
\end{equation}%
After manipulating Eq.s (51) and (52) one finds%
\begin{equation}
\nabla ^{2}\phi =f\phi ^{\prime \prime }+f^{\prime }\phi ^{\prime }=V\left(
\phi \right) 
\end{equation}%
and 
\begin{equation}
\tilde{R}=-f^{\prime \prime }=-V^{\prime }\left( \phi \right) ,\text{ }
\end{equation}%
in which a prime means derivative with respect to the argument. Our dilaton
ansatz 
\begin{equation}
\phi =\tilde{x}
\end{equation}%
admits%
\begin{equation}
f^{\prime }=V\left( \phi \right) ,
\end{equation}%
such that 
\begin{equation}
f\left( \phi \right) =J\left( \phi \right) -\mathcal{C}
\end{equation}%
in which 
\begin{equation}
J\left( \phi \right) =\int V\left( \phi \right) d\phi =\frac{\phi ^{2}}{\ell
^{2}}-\frac{2\pi \left( 1-2k\right) ^{2}q^{2k}}{2^{k+1}\left( k-1\right) }%
\left( \frac{1}{\phi ^{\frac{2\left( 1-k\right) }{2k-1}}}-\frac{1}{\phi
_{0}^{\frac{2\left( 1-k\right) }{2k-1}}}\right) .
\end{equation}%
Herein $\phi _{0}$ is a reference potential, $\tilde{\Lambda}=-\frac{1}{\ell
^{2}}$ and $\mathcal{C}$ represents the ADM mass of the EPM black hole \cite%
{25,26,27}. Also the line element (54) becomes%
\begin{equation}
ds^{2}=-f\left( \phi \right) dt^{2}+\frac{d\phi ^{2}}{f\left( \phi \right) }.
\end{equation}%
As it was introduced in \cite{25,26,27}, the extremal value of $\phi _{+}$
is obtained from $V\left( \phi _{+}=\phi _{e}\right) =0,$ which yields%
\begin{equation}
\phi _{e}^{\frac{2k}{2k-1}}=\frac{2\pi q^{2k}\ell ^{2}}{2^{k+1}}\left(
2k-1\right) .
\end{equation}%
This implies that the extremal mass (from (60)) is given by 
\begin{equation}
M_{e}=J\left( \phi _{e}\right) ,
\end{equation}%
in which for $M\geq M_{e}$ the metric function admits at least one horizon $%
\phi _{+}$ that indicates the outer horizon. The Hawking temperature at the
outer horizon, the heat capacity and free energy are given respectively by%
\begin{equation}
T_{H}=\frac{V\left( \phi _{+}\right) }{4\pi }=\frac{\phi _{+}}{4\pi }\left( 
\frac{2}{\ell ^{2}}+2\pi \left( \frac{1}{2^{k}}\frac{q^{2k}}{\phi _{+}^{%
\frac{2k}{2k-1}}}\right) \left( 1-2k\right) \right) ,
\end{equation}%
\begin{equation}
C_{q}\left( \phi _{+}\right) =4\pi \left( \frac{V\left( \phi _{+}\right) }{%
V^{\prime }\left( \phi _{+}\right) }\right) =\frac{4\pi \phi _{+}\left( 
\frac{2}{\ell ^{2}}+2\pi \left( \frac{1}{2^{k}}\frac{q^{2k}}{\phi _{+}^{%
\frac{2k}{2k-1}}}\right) \left( 1-2k\right) \right) }{\frac{2}{\ell ^{2}}%
+2\pi \left( \frac{1}{2^{k}}\frac{q^{2k}}{\phi ^{\frac{2k}{2k-1}}}\right) }
\end{equation}%
and%
\begin{equation}
F\left( \phi _{+}\right) =\frac{\phi _{+}^{2}}{\ell ^{2}}-\frac{2\pi \left(
1-2k\right) ^{2}q^{2k}}{2^{k+1}\left( k-1\right) }\left( \frac{1}{\phi _{+}^{%
\frac{2\left( 1-k\right) }{2k-1}}}-\frac{1}{\phi _{0}^{\frac{2\left(
1-k\right) }{2k-1}}}\right) -J\left( \phi _{e}\right) -\phi _{+}^{2}\left( 
\frac{2}{\ell ^{2}}+2\pi \left( \frac{1}{2^{k}}\frac{q^{2k}}{\phi _{+}^{%
\frac{2k}{2k-1}}}\right) \left( 1-2k\right) \right) .
\end{equation}

In summary, in this section the thermodynamic analysis of the EPM black hole
is investigated by two entirely different methods. Our analysis reveals that
by rescaling the constant $q$ we recover the results obtained in so (40)
that two different approaches for thermodynamic stability are in agreement.

\section{CONCLUSION}

In this study, the most general solution in $(2+1)-$dimensional spacetime in
EPM theory, without imposing the traceless condition on the energy -
momentum tensor is derived. The obtained solutions describe black holes
sourced by the power Maxwell fields. From physics standpoint and in analogy
with the self - interacting scalar fields, $k$ can be interpreted as the
measure of self - interaction that electromagnetic field undergoes. As such,
it alters much of physics and, in particular, the black hole / singularity
formations. We have shown with particular examples that the power parameter $%
k$ has a significant effect on the physical interpretation of the obtained
solutions. For specific values of parameter $k,$ it is possible to obtain
asymptotically flat ( with $\Lambda =0$) or non - asymptotically flat
solutions. As it has been shown in the Appendix, for the choice $\frac{2}{3}%
<k<1,$ all energy conditions ( WEC, SEC and DEC) are satisfied, as well as
the causality condition. The character of the singularity at $r=0$ is
timelike, since a new coordinate defined by $r_{\ast }=\int \frac{dr}{f}$ \
is finite as $r\rightarrow 0$. In solutions admitting black holes, this
timelike singularity is covered by horizon(s). But, in some cases it remains
naked and violates the cosmic censorship hypothesis. It becomes worthful
therefore to investigate the structure of this singularity in quantum
mechanical point of view. For $\frac{2\left( k-1\right) }{2k-1}>0$ the
resulting spacetime geometry is very similar to the BTZ black hole whose
quantum singularity structure is investigated in \cite{28} by quantum test
particles obeying the Klein-Gordon and Dirac equations. The results reported
in \cite{28} are; for massive scalar fields the spacetime is quantum
singular but for massless scalar bosons and for fermions, the spacetime is
quantum regular. On the other hand, naked singularity that occurs for $\frac{%
2\left( k-1\right) }{2k-1}<0$ is structurally similar to the solution given
in \cite{12}. The quantum nature of this singularity is recently
investigated in \cite{29} with the test particles obeying the Klein-Gordon
and Dirac equations. It was shown that the spacetime is quantum singular for
massless scalar particles obeying Klein-Gordon equation but quantum regular
for fermions obeying Dirac equation. Therefore, these results are also
applicable to the solutions presented in this study. Thermodynamic
quantities such as Hawking temperature, entropy and specific heat capacity
are also calculated by two different methods in which we obtain the same
result for stability. Magnetically charged non-black hole EPM solutions are
considered in a recent study \cite{30} in which singularities, both
classically and quantum mechanically are investigated thoroughly. The fact
that by employing NED in $2+1-$dimensions one can construct regular black
holes through cutting and pasting method has also been shown in a separate
study \cite{31}. Finally, we note that by choosing $k\neq 1$ in the flat
spacetime electrodynamics, we avoid the logarithmic potential, once and for
all.

When a matter field couples to any system, energy conditions must be
satisfied for physically acceptable solutions. This is achieved by following
the steps as given in \cite{32,33}.$T_{\ \nu }^{\mu }=\frac{1}{2}\left\vert 
\mathcal{F}\right\vert ^{k}$diag$\left( \xi ,\xi ,-1\right) $

\subsection{Weak Energy Condition (WEC)}

The energy - momentum tensor is given in Eq.(14) implies

\begin{equation*}
\rho =-T_{t}^{t}=\frac{1}{2}\left\vert \mathcal{F}\right\vert ^{k}\xi ,\text{
\ \ \ \ \ \ \ \ }p_{r}=T_{r}^{r}=\frac{1}{2}\left\vert \mathcal{F}%
\right\vert ^{k}\xi ,\text{ \ \ \ \ \ \ \ \ }p_{\theta }=-\frac{1}{2}%
\left\vert \mathcal{F}\right\vert ^{k},
\end{equation*}%
in which $\rho $ is the energy density and $p_{i}$ are the principal
pressures.

The WEC states that,

\begin{equation}
\rho \geq 0\text{ \ \ \ \ \ \ \ \ \ \ \ and \ \ \ \ \ \ \ \ }\rho +p_{i}\geq
0\text{ \ \ \ \ \ }(i=1,2)  \tag{A1}
\end{equation}%
which imposes $k>\frac{1}{2}.$

\subsection{Strong Energy Condition (SEC)}

This condition states that;

\begin{equation}
\rho +\dsum\limits_{i=1}^{2}p_{i}\geq 0\text{ \ \ \ \ \ \ and \ \ \ \ \ \ }%
\rho +p_{i}\geq 0,  \tag{A2}
\end{equation}%
which yields $k\geq 0.$ The SEC together with the WEC constraint the
parameter $k$ to \ $k>\frac{1}{2}.$

\subsection{Dominant Energy Condition (DEC)}

DEC states that $p_{eff}\geq 0$ in which

\begin{equation}
p_{eff}=\frac{1}{2}\dsum\limits_{i=1}^{2}T_{i}^{i}.  \tag{A3}
\end{equation}%
and this gives the constraint $k\leq 1.$ One can show that DEC, together
with SEC and WEC, imposes $\frac{1}{2}<k\leq 1.$ \ 

\subsection{Causality Condition (CC)}

Beside the energy conditions, one can impose the causality condition (CC)

\begin{equation}
0\leq \frac{p_{eff}}{\rho }<1,  \tag{A4}
\end{equation}%
which implies $\frac{2}{3}<k\leq 1.$ Therefore if CC is imposed, naturally
all other conditions are satisfied.

\end{document}